# OPTIMIZING SANDWICH PANELS WITH GRADED TUBULAR CELL CORE FOR ENHANCED SOUND TRANSMISSION LOSS

M. M. Keleshteri[1, 2] and J. Jelovica[3]

## Abstract

In this research a theoretical approach is proposed to evaluate sound transmission loss (STL) of a sandwich panel with tubular cells core (TCC) in the in-plane orientation. Afterwards, the core geometric parameters are optimized by implementing the Non-dominated Sorting Genetic Algorithm II (NSGA-II) to maximize the STL through the sandwich panel while minimizing the total mass and height of the panel. The sound transmission response of the sandwich panel is parameterized in terms of four independent geometric parameters of the core, which are: number of tubular cells in the length and height directions of the core, thickness of the tubular cell's wall and radii of the tubular cells. It is assumed that every tubular cell has a similar thickness while their radius in each row differs. Using the Abaqus scripting interface, a structural acoustic finite element model of the sandwich panel with tubular cell core is generated based on the specified parametric parameters. The study designs are obtained within a frequency range of 50Hz to 350Hz, which includes the first few resonance regions of the sandwich panel. In order to decrease the computational cost, a machine learning technique based on gradient boosting decision tress (GBDT) is employed to predict the STL of the sandwich panel. Findings reveal that the GBDT model is successfully trained. Data comparison shows that the discrepancies between the GBRT model and the FE model are confined to a margin of -3.09% to 2.7% in 95% of instances. The training is comprised of 500 examples, and an extra 150 samples are employed for testing to guarantee the reliability of the outcomes. The optimized panel using tubular cells demonstrates 52% and 23% reduction in the mass and height of the panel, respectively, when compared to the reference panel with the same area under the STL vs. frequency curve.

[1] Department of Mechanical Engineering, The University of British Columbia, Vancouver, BC, Canada.
[2] Corresponding Author, Email: m.keleshteri@ubc.ca
[3] Departments of Mechanical Engineering and Civil Engineering, The University of British Columbia, Vancouver, BC, Canada.

## 1. Introduction

The unique structural design of sandwich panels contributes to their popularity in acoustic applications due to the exceptional sound insulation and absorption properties they offer. The combination of two thin and stiff faces bonded to a low-density core creates a layered structure that effectively reduces sound transmission [1]. The faces can carry in-plane loads and bending moments, while the core, with its lower stiffness compared to the skins, resists transverse shear loads. This structural arrangement allows sandwich panels to effectively dampen and absorb sound waves, resulting in excellent sound insulation performance. Due to the exceptional sound insulation and absorption properties of sandwich panels, numerous studies have focused on the acoustic analysis of these panels, exploring various configurations [1]–[29]. Researchers have investigated different aspects such as the effects of core materials, skin thickness, core geometry, and bonding techniques on the acoustic performance of sandwich panels. By analyzing these factors, researchers aim to understand and optimize the sound transmission and absorption characteristics of sandwich panels in different applications. These studies provide valuable insights into design of sandwich panels with enhanced acoustic performance, enabling their effective use in diverse fields such as architecture, aerospace, automotive, and industrial settings [1], [9], [30]. A concise overview of the existing literature is provided below.

Fu et al. [4] investigated the sound transmission loss (STL) properties of sandwich plates with various truss cores, including pyramidal core, tetrahedral core, and 3D-kagome core. Their research indicated that the sandwich plate with a pyramidal core demonstrates a lower resonance frequency, while also delivering superior sound insulation performance across a broad frequency range. Wang et al. [9] introduced a numerical approach based on the higher-order sandwich plate theory to predict the sound transmission loss (STL) of sandwich panels using statistical energy analysis (SEA). Their approach considers both antisymmetric and symmetric (dilatational) motions, treating these motions as independent factors in the analysis. In their research, Tao et al. [12] examined the sound radiation characteristics of a truss core sandwich structure under thermal conditions. They analyzed the influence of various important factors such as the radius, inclination angle, and height of the truss core, as well as the temperature field. The study revealed that the peaks in the sound power level (SPL) curve consistently corresponded to the natural frequencies. Furthermore, they observed that increasing thermal loads resulted in a shift of SPL peaks and dips towards lower frequencies. This shift was primarily attributed to changes in material properties and internal stress caused by the temperature rise, leading to a reduction in structural stiffness. Wen et al. [18] conducted a study on the sound transmission loss of sandwich panels with a closed octahedral core (SPCOC). They discovered that, when compared to corrugated sandwich panels (CSP), SPCOC exhibits remarkable sound insulation performance at frequencies below 1600 Hz. Interestingly, despite having a unit cell mass that is 14.181% lower than that of CSP, SPCOC demonstrated an increase in sound insulation properties ranging from 4.61% to 20.62% within the frequency range of 200-1600 Hz. Arunkumar et al. [22] examined the impact of core geometry, including honeycomb, truss, z-shaped, and foam cores, on the vibroacoustic response of sandwich structures. Their findings indicate that, for honeycomb core panels, reducing core height and increasing face sheet thickness can enhance acoustic comfort. They also observed that vibration and acoustic response in honeycomb-core panels are not significantly influenced by cell size. The study showed that triangular cores offer better acoustic comfort in truss core panels compared to other core types.

Sandwich panels featuring a cellular core possess advantageous characteristics for acoustic applications due to their distinct geometric configuration [6], [25], [31]–[33]. However, a review of existing literature has revealed that there has been a lack of research on the vibroacoustic behavior of sandwich panels with tubular cell core (TCC). The tubular cells allow for high sound absorption, enabling sound waves to enter and dissipate within the cells. Additionally, the acoustic properties of the panel can be adjusted by varying the thickness and radius of the tubular cells, allowing for customization to specific applications, such as in the aerospace and automotive industries. However, optimizing the acoustic performance of TCC sandwich panels can be challenging due to the complex nature of the core structure. In this context, the application of machine learning approaches holds promise in predicting and optimizing the sound transmission loss of these panels, offering a potential solution to overcome these challenges.

This study introduces a machine learning model to evaluate the sound transmission loss (STL) of sandwich panels featuring a tubular cell core. Subsequently, the geometric parameters of the core are optimized using the NSGA-II algorithm to maximize the STL while simultaneously minimizing the total mass and height of the panel. The results demonstrate the effectiveness of the proposed machine learning approach in identifying the optimal combination of geometric parameters to achieve maximum sound transmission loss. By utilizing machine learning technique to evaluate and optimize the sound transmission loss, this research offers a data-driven approach that can significantly streamline the design process of TCC sandwich panels. The ability to identify the most effective geometric parameters allows for enhanced acoustic performance while considering practical constraints such as total mass and height. This research contributes to the advancement of acoustic engineering and provides valuable insights for industries aiming to develop high-performance sound insulation solutions with improved efficiency and accuracy.

## 2. TCC Sandwich Panels

The sandwich panels under analysis consist of three layers. These panels are composed of a tubular cell core that is sandwiched between two identical face sheets. The face sheets and the core are both made of aluminum, with a modulus of elasticity, $E$, of 71.9 $GPa$, a Poisson's ratio, $v$, of 0.3, and a density, $\rho$, of 2700 $kg/m^3$. The face sheets have a length, $L$, of 1 $m$ and a thickness, $t_f$, of 2.5 $mm$. The core's height, $H$, is determined by the number of tubular cells, $M$, in the vertical $y$ direction. The wall thickness of the tubular cells, $t_c$, is assumed to be constant. However, while the radius of the tubular cells, $R$, remains constant within a row, it varies in the vertical direction, as shown in Figure 1. In addition, the spacing between the tubular cells in the core is denoted as $S$ and can be calculated using:

$$S = \frac{L}{N} \tag{1}$$

where, $N$ is the number of tubular cells in the length direction ($x$-direction).

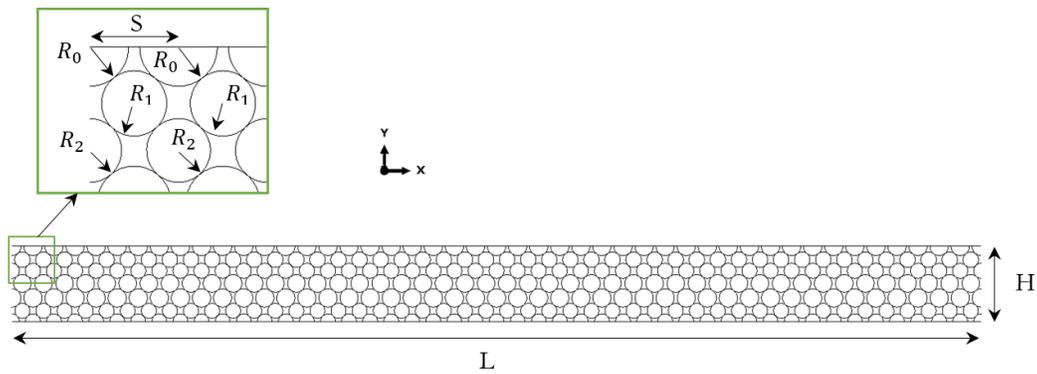

**Figure 1:** Illustrating the structural geometry of a TCC sandwich panel.

## 3. Structural Acoustics

This study is concerned with sound transmission between two distinct types of media: a solid structure, which is a complex sandwich panel, and a fluid medium, in particular air. Each region of the solid or fluid medium has its own unique material properties and must be discretized into finite elements with suitable size and aspect ratio. Additionally, the interaction between different regions must be precisely defined at the interfaces to accurately capture the sound transmission behavior. The motion of the structure is governed by linear elastic behavior, assuming that it follows the principles of linear elasticity [34]:

$$\nabla . \boldsymbol{\sigma} + \boldsymbol{F} = \rho \ddot{\boldsymbol{u}} \tag{2}$$

$$\boldsymbol{\sigma} = \boldsymbol{C} : \boldsymbol{\epsilon} \tag{3}$$

$$\boldsymbol{\epsilon} = \frac{1}{2} [\nabla \boldsymbol{u} + (\nabla \boldsymbol{u})^T] \tag{4}$$

where $\boldsymbol{\sigma}$ is the Cauchy stress tensor, $\boldsymbol{F}$ is the external force per unit volume, $\nabla$ is the gradient operator in the appropriate co-ordinate frame, $\boldsymbol{\epsilon}$ is the infinitesimal strain tensor and $\boldsymbol{C}$ is the stiffness tensor. Simplifications can be applied to adapt the equations to the specific characteristics and behavior of the geometries under investigation, such as shells or beams.

The Helmholtz equation is a partial differential equation that describes the behavior of wave propagation in a given medium. In the context of acoustics, it is used to analyze the propagation of sound waves. The general form of the Helmholtz equation is [35]:

$$\nabla^2 \psi + k^2 \psi = 0 \tag{5}$$

where $\nabla^2$ represents the Laplacian operator, $\psi$ is the scalar field representing the sound pressure or acoustic potential, $k$ is the wave number (equal to $2\pi$ divided by the wavelength). The equation is typically solved for a specific frequency. The way the fluid and structural domains interact with each other is determined by [35]:

$$\frac{\partial P}{\partial n} + \rho \frac{\partial^2 u_n}{\partial t^2} = 0 \qquad (6)$$

The displacement of the solid surface in the direction normal to the structure's surface is represented by $u_n$, and the pressure exerted on the panel's face sheet on the incident side is denoted as $P$. By dividing the fluid and solid regions into a finite number of elements, it becomes feasible to discretize the material equations and establish a system of equations where the unknowns are the degrees of freedom for each element. Solving this system of equations for a significant number of elements, considering appropriate aspect ratios, allows us to determine the values of the degrees of freedom at each point within the domain. To expedite the computation of sound transmission loss (STL) calculations, the commercially available finite element package ABAQUS is employed in this study. It is worth noting that alternative techniques like boundary element methods and spectral methods can also be utilized for this purpose [19].

### 3.1. FEM Approach

In this research, the structural-acoustic analysis utilizes a finite element (FE) model that represents an in-plane tubular cells core sandwich panel with an air domain, as depicted in Figure 2. The FE modeling is carried out using the commercially available software ABAQUS. The decision to adopt the in-plane configuration for the sandwich panel in the model is based on its higher sound transmission loss response compared to the out-of-plane configuration, as mentioned in Ref. [3]. The model space consists of two distinct regions: the incident side positioned below the sandwich panel and the transmitted side located above it. The transmitted side represents the air domain. To simulate a plane incident wave, a uniform harmonic pressure with a unit amplitude is applied perpendicular to the lower face sheet of the sandwich panel on the incident side. The STL response of a structure to pressure waves is influenced by the angle at which the wave hits it. Ref. [36] indicates that the structure exhibits the highest response when the wave hits it head-on at a 0-degree angle. Therefore, the current study employs a plane wave with a 0-degree incident angle. Under the applied pressure, the plate undergoes vibratory motion, resulting in sound generation on the transmitted side. Previous research [37] suggests that replacing the vacuum in the open spaces of the core with air does not significantly affect the accuracy of the STL response. Consequently, for computational efficiency, the open cavities within the core of the sandwich panel are assumed to be filled with vacuum. The ends of the sandwich panel are subjected to "pinned" type boundary conditions, which restrict translation in the $x$ and $y$ directions while allowing free rotation, as illustrated in Figure 2.

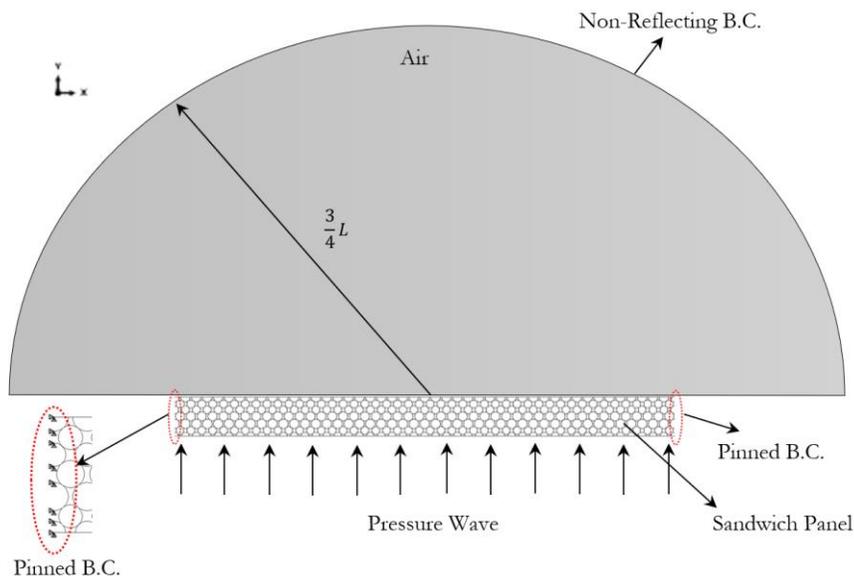

**Figure 2:** Visual representation of the finite element model for structural-acoustic analysis

In the structural-acoustic model, there is a designated air region with mass density of 1.2 $kg/m^3$ and bulk modulus of 141.2 $N/m^2$ that usually takes on a semi-circular shape. The diameter of this semi-circle is 1.5 times the length of the sandwich plate. Although the air region has a finite size, the goal is to simulate the behavior as if it were an infinitely large air region. To achieve this, a non-reflecting boundary condition is applied specifically along the semicircular edge of the air region. This boundary condition is designed to prevent any sound waves transmitted through the air region from reflecting back from the semicircular edge. Instead, the condition ensures that these waves continue to propagate towards infinity without any reflections occurring. By implementing this non-reflecting boundary condition, the model can effectively mimic the behavior of an infinitely large air region, allowing for a more accurate representation of how sound waves travel through the system. This approach helps to capture the realistic behavior of sound transmission and ensures that the simulation results are not influenced by artificial reflections at the semicircular edge of the air region.

The tubular cell core and face sheets have been discretized in the model using quadratic Timoshenko beam elements with three nodes (specifically B22 elements in ABAQUS). On the other hand, the air domain has been discretized using linear 2D acoustic triangular elements with three nodes (specifically AC2D3 elements in ABAQUS). The mesh for the air domain exhibits non-uniformity, as illustrated in Figure 3. To ensure accurate representation of the pressure changes caused by the vibrating panel, the mesh size is finer in the vicinity of the sandwich panel compared to the semicircular geometry. This allows for a more precise capture of the pressure variations. The connection between the air and the sandwich panel is established using "tie constraints" in ABAQUS.

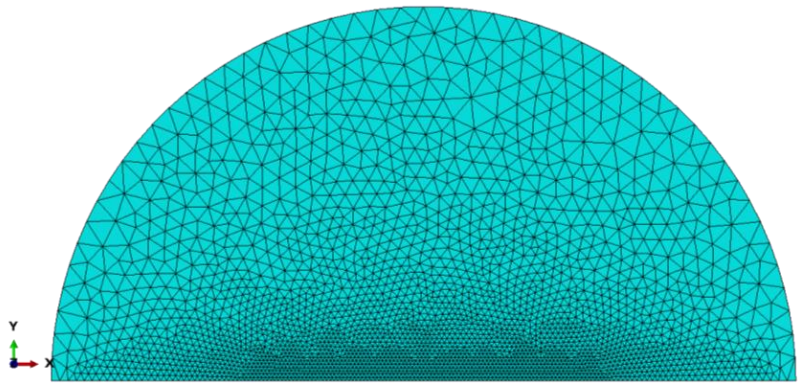

**Figure 3:** The mesh used for the air domain is designed with a greater density towards the edge that is in contact with the sandwich panel.

where $c$ represents the wave speed and $f$ is the frequency in $Hz$. In this study, the highest frequency of interest is $f_{max} = 350\ Hz$. For bending waves, the wave speed is frequency-dependent and can be approximated by [38]:

$$c = \sqrt[4]{\left(\frac{E(2\pi tf)^2}{12\rho}\right)} \qquad (8)$$

where $E$ is the Young's modulus, $t$ is the thickness, and $\rho$ is the density of the material. At $f_{max}$, the wave speed for the properties of the reference design is $90.5\ m/s$, which corresponds to a minimum wavelength of $0.26\ m$. The largest beam element length in the model is $0.005\ m$, which corresponds to 52 elements per wavelength. This indicates a highly refined mesh, as there are a significant number of elements per wavelength.

### 3.2. STL Calculation

The sound transmission loss serves as a widely used performance measure to evaluate the sound insulation properties of a material [39], [40]. It represents the ratio between the incident sound pressure and the transmitted sound pressure across the surface of a partition or panel. The sound transmission loss (STL) of a panel is dependent on the frequency and is influenced by both the material composition and the geometry of the panel. The primary factor determining the STL varies depending on the specific frequency range under consideration. The calculation of STL for the sandwich panel at a specific frequency involves determining the ratio between the sound pressure of the incident and transmitted sound, as described in reference [39].

$$STL = 10 log\left|\frac{P_i^2}{P_t^2}\right| \qquad (9)$$

where $P_i$ and $P_t$ are the root mean square value of pressure on the incident side and on the transmitted side with the unit of $(N/m^2)$, and $STL$ is the sound transmission loss $(dB)$. These values are computed numerically from:

$$P_i^2 = \boldsymbol{p}_i \cdot \boldsymbol{p}_i \quad , \quad P_t^2 = \boldsymbol{p}_t \cdot \boldsymbol{p}_t \tag{10}$$

where

$$\boldsymbol{p} = \langle p_1, p_2, \ldots, p_n \rangle \tag{11}$$

is a vector of the pressure values at the air FE nodes along the face sheets of the panel. In the steady state analysis conducted with ABAQUS, a history output is defined for the acoustic pressure of the air nodes directly in contact with the transmitted (top) side of the sandwich panel. The acoustic pressure results are expressed as complex numbers, and their magnitudes are recorded. As the loading is specified as a unit pressure wave, the incident pressure on all nodes on the incident side of the panel is set to one. These results are then utilized to calculate the sound pressure transmission loss of each panel using Equation (4).

## 4. Training a ML Model

In this research a gradient boosting decision tree (GBDT) is utilized as machine learning approach to determine the sound transmission loss of TCC sandwich panels. The GBDT is a versatile machine learning algorithm utilized for regression and classification tasks that has a wide range of commercial and academic applications [41]. It operates as an ensemble learning technique by combining multiple weak models, known as decision trees, to form a robust predictive model. In the field of machine learning, ensemble learning refers to the utilization of multiple learning algorithms in order to achieve superior predictive performance compared to what could be accomplished with any single learning algorithm in isolation [42].

The concept behind GBDT involves iteratively training new models that aim to minimize the errors of the previous models [43], [44]. In each iteration, the new model is trained to predict the discrepancies between the previous model's estimated values and the actual values. These discrepancies, also called residuals, serve as the target for the new model. By updating the weights of the training data based on these residuals, the subsequent model can focus on areas where the previous model exhibited poor performance. This iterative process continues until the desired level of accuracy is attained. An overview of gradient boosting decision tree architecture is shown in Figure 4.

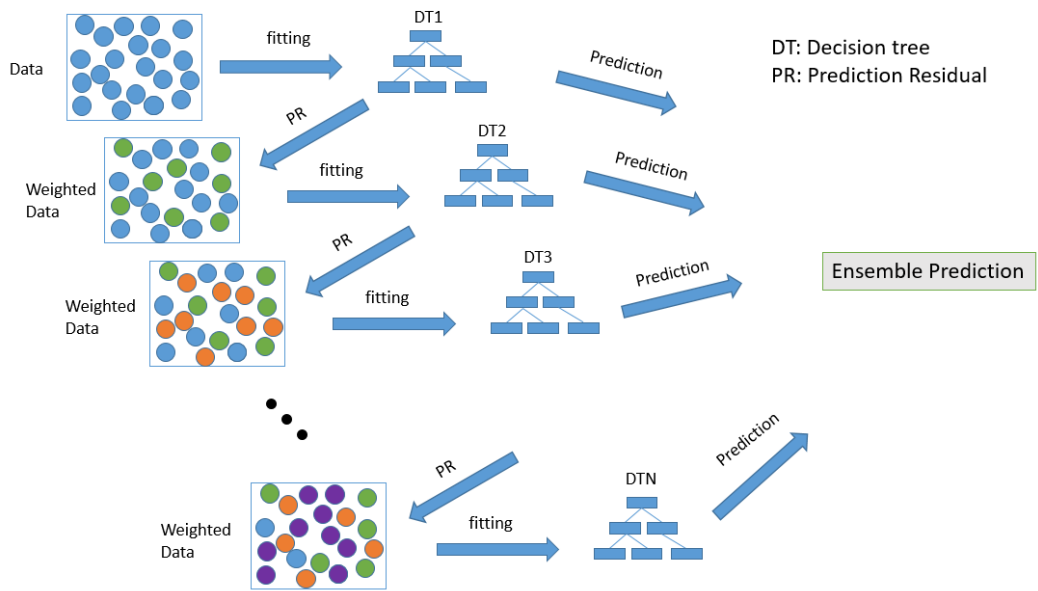

**Figure 4:** An Overview of Gradient Boosting Decision Tree Architecture

The mathematical representation of the GBDT is provided in Equation 12. The objective of the training process is to discover an estimated function $F(x)$ that minimizes the expected value of a loss function $L(y, f(x))$. This is accomplished by combining a series of weak learners, represented as functions $h_i(x)$, using a weighted sum. These weak learners are typically small decision trees of fixed size. The additive model introduced by gradient boosting can be represented as shown in Equation

12, where M denotes the number of classifiers utilized and $\gamma_m$ represents the weight associated with each classifier.

$$F(x) = \sum_{m=1}^{M} \gamma_m h_m(x) \tag{12}$$

In gradient boosting, the aforementioned additive model is constructed in a forward stagewise manner. The iteration process can be described as follows:

$$F_m(x) = F_{m-1}(x) + \gamma_m h_m(x) \tag{13}$$

where $m$ represents the $m^{th}$ iteration. During the $m^{th}$ iteration, the regression tree $h_m(x)$ is fitted using the negative gradient in the following manner [43]:

$$r_i = -\left[\frac{\partial L(y_i, F(x_i))}{\partial F(x_i)}\right]_{F(x)=F_{m-1}(x)}, \quad i = 1, \ldots, n \tag{14}$$

where $L$ represents the chosen loss function. $n$ denotes the number of data samples, and the step length is selected to minimize the overall loss, as implied by [43], [45]:

$$\gamma_m = arg_\gamma \min \sum_{i=1}^{n} L(y_i, F_{m-1}(x_i) + \gamma h_m(x_i)) \tag{15}$$

By incorporating the shrinkage parameter, $v$, to regulate the learning rate, Equation 13 is modified as follows [43]:

$$F_m(x) = F_{m-1}(x) + v\gamma_m h_m(x) \tag{16}$$

More comprehensive information regarding gradient boosting regression can be found in references [43]–[46]. The implementation of gradient boosting regression in this paper is performed using the scikit-learn library [47].

In this research, the GBDT is applied to predict the area under the Sound Transmission Loss (STL) versus frequency curve for a sandwich panel. To quantify the total STL response of the sandwich panel within the given frequency range, the study adopted the Area Under the Curve (AUC) as a scalar measurement. AUC summarizes the overall performance of the panel in terms of sound transmission loss. The calculation of AUC involves integrating the STL versus frequency curve between the lower frequency bound, $f_1$, and the higher frequency bound, $f_2$, capturing the cumulative effect of sound attenuation across the entire frequency range. The calculation for this measurement involves:

$$AUC = \int_{f_1}^{f_2} (STL) df \qquad (17)$$

By utilizing GBDT and AUC, this study aims to provide accurate predictions of the sound insulation characteristics of the sandwich panel, supporting its optimization and assessment in practical applications.

In order to collect data from the Finite Element (FE) simulation, the construction of the sandwich panel involves establishing constraints on the design variables. These variables include $N$, $M$, $t_c$, and $R_i$ (where $i$ ranges from 0 to $M - 1$). Upper and lower boundaries are defined to specify the acceptable ranges for these design variables as:

- $2 \leq M \leq 10$
- $10 \leq N \leq 100$
- $0.3 \leq \frac{R_i}{S} \leq 0.5$ ; $i = 1, 2, \ldots, M$
- $0.05 \leq \frac{t_c}{R_{min}} \leq 0.3$

These boundaries serve as guidelines to ensure that the simulation captures a meaningful range of values for each variable, facilitating a comprehensive analysis of the sandwich panel's behavior and performance. By imposing such limits on the design variables, the FE simulation can generate data that encompasses various configurations of the sandwich panel, enabling a comprehensive evaluation and exploration of its characteristics.

Given the bounds imposed on the design variables, a total of 650 samples were generated for this study. Among these samples, 500 were allocated for training the machine learning (ML) platform, while the remaining samples were reserved for testing the trained model. To optimize machine learning algorithm's performance, specific settings were carefully chosen for the GBDT algorithm. The learning rate, set to 0.15, controls the step size during each iteration of the gradient descent process. Smaller values result in slower convergence but higher accuracy, whereas larger values lead to faster convergence but lower accuracy. Similarly, the number of estimators, which was determined as 450 in this research, defines the number of decision trees used in the ensemble model. Larger values generally yield better model performance but increase the risk of overfitting the data. Therefore, the values for the learning rate and the number of estimators were selected through a series of iterative experiments to strike a balance between model accuracy and computational efficiency.

## 5. Design Optimization for Maximum STL

Different from the single-objective optimization problem that yields only one optimal solution, the solution to a multi-objective optimization problem is a set of points called the Pareto optimal set, which represents a range of trade-off solutions between conflicting objectives. Numerous techniques have been proposed in the literature [48] to obtain these Pareto-optimal solutions, with multi-objective evolutionary algorithms (MOEAs) such as Non-dominated Sorting Genetic Algorithm II (NSGA-II) [49], Strength Pareto Evolutionary Algorithm II (SPEA-II) [50], and multi-objective evolutionary algorithm based on decomposition (MOEA/D) [51] receiving significant attention due to their effectiveness and ease of implementation. Among these algorithms, NSGA-II stands out as a particularly powerful method. It is an enhanced version of NSGA [52] developed from the widely-used genetic algorithm and the concept of non-dominated sorting introduced by Goldberg [53]. Over the past decade, NSGA-II has undergone improvements and has been extensively applied in the design optimization of various problems [54]–[56], including optimizing vibroacoustic behavior of sandwich panels [57], [58]. In this research, the NSGA-II algorithm is utilized to perform a global multi-objective optimization. The application of NSGA-II to the optimization of sandwich panel designs for maximum STL offers several advantages. Firstly, it allows for the simultaneous consideration of multiple design variables, such as core geometry and panel dimensions. Secondly, NSGA-II enables the exploration of a wide range of design possibilities, searching for the most favorable trade-offs between enhancing STL, minimizing panel mass, and reducing overall panel height. The brief description of the algorithm is presented as follows:

- Generate an initial population, $P_0$, containing $N_N$ individuals.
- Create an offspring population, $Q_t$, through binary tournament selection based on a crowding-comparison operator, as well as crossover and mutation operations performed on the parent population, $P_t$. The subscript '$t$' indicates the generation number. The offspring population, $Q_t$, and the parent population, $P_t$, are then combined to form the entire population, $R_t$.
- Apply a fast non-dominated sorting approach to the entire population, $R_t$, to identify distinct non-dominated fronts for the objective functions $F_1$, $F_2$, and so on.
- Construct a new parent population, $P_{t+1}$, consisting of $N_N$ individuals selected from the obtained fronts, $F_i$.
- Repeat the process until the maximum number of iterations is reached.

Variables are represented as continuous numbers. Simulated binary crossover (SBX) operator and polynomial mutation are used, with distribution indexes for crossover $\eta_c$ and mutation $\eta_m$ both

20, according to [49]. Probability of crossover is 0.9 and probability of mutation is $1/nv$, where $nv$ is the number of design variables. The algorithm is run until convergence, with a population size of 100.

More comprehensive information regarding algorithm's procedure can be found in [49]. The implementation of the NSGA-II in this paper is performed using the jMetal library [59] with the goal of maximizing the STL while minimizing the total mass and height of the panel considering following conditions:

- **Design Parameters:** The overall length of the sandwich panel, thickness of the face sheets, angle of the incident pressure wave, boundary conditions and material properties of the core, face sheets and the air are maintained constant.

- **Independent Design Variables:** These variables include the number of tubular cells in the thickness direction of the panel $(2 \leq M \leq 10)$, the number of tubular cells in the direction of the panel's length $(10 \leq N \leq 100)$, the thickness of the wall of the tubular cells $(0.015 \leq t_c/S \leq 0.09)$ and the radius of the tubular cells $(0.3 \leq R_i/S \leq 0.5)$.

- **Derived Variables:** These variables include the height of the core, $H$, spacing of the core, $S$, mass of the panel, $M$, and the area under the STL versus frequency curve, $AUC$, which can be defined as a function of dependent variables as:

$$H = F(R_i, L, N, t_c) \tag{18}$$

$$S = F(L, N) \tag{19}$$

$$M = F(G, MT) \tag{20}$$

$$AUC = F(G, MT, P, B, f) \tag{21}$$

Where $G, MT, P, B$ and $f$ represent the geometrical parameters, material properties, loading conditions, boundary conditions and range of the desired frequency, respectively.

## 6. Results and Discussions

In order to verify the accuracy of the results, first a mesh study was conducted to determine the mesh convergence for the STL response. The results of the mesh convergence analysis for the reference design within the desired frequency range are depicted in Figure 5. The analysis commenced with an initial configuration referred to as "Mesh 1," which consisted of approximately 5,000 elements for the core, 100 elements for the face sheets, and 3,000 elements for the air domain. Following that, the number of elements was increased by multiplying the mesh density in Mesh 1 with an increasing integer until the percentage difference in the root-mean-square value of STL between two consecutive meshes was determined to be less than one percent. This indicates successful mesh convergence. Consequently, "Mesh 4" was selected, consisting of approximately 20,000 elements for the core, 400 elements for the face sheets, and 12,000 elements for the air domain, for the remaining calculations in this study.

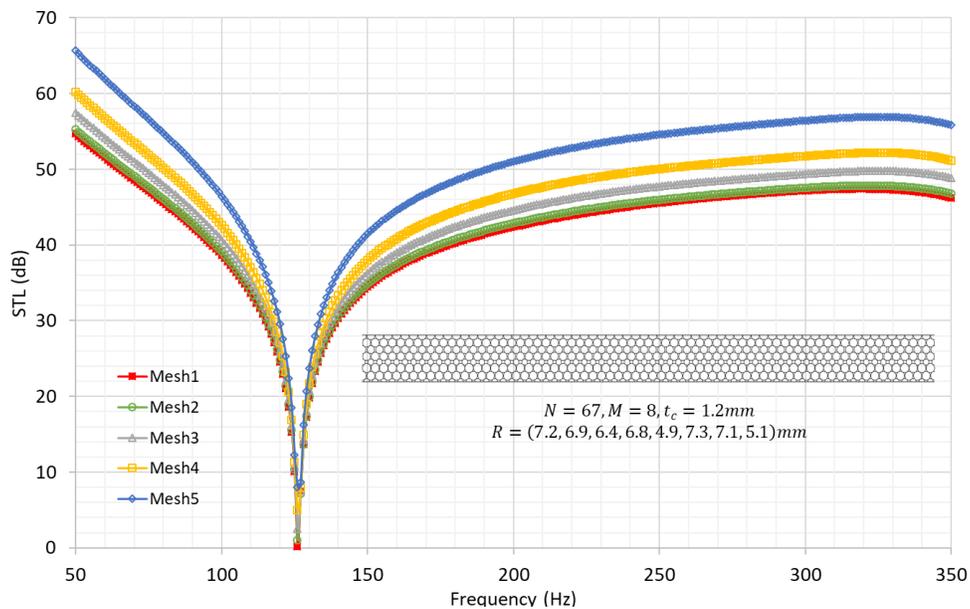

**Figure 5:** Mesh convergence study

Next, in order to demonstrate the accuracy of the present finite element solution procedure, the STL versus frequency graph of a honeycomb sandwich panel were compared with that of the same model from Griese et al. [20] and Galgalikar and Thompson [14] in Figure 6. The referenced papers provide the specific geometric and material parameters used in the analysis. The choice of a honeycomb sandwich panel was made because the STL response of TCC sandwich panels has not been investigated before. Therefore, by examining the response of a honeycomb sandwich panel, a reference point could be established to assess the validity of the current approach. The outcome of the comparison demonstrated a highly favorable agreement between the present method and the previous studies. This suggests that the current finite element solution procedure accurately captures the behavior of the honeycomb sandwich panel, as indicated by the close alignment of the STL versus frequency graph. The observed consistency between the approaches reinforces the reliability and robustness of the current method in analyzing the STL characteristics of sandwich panels.

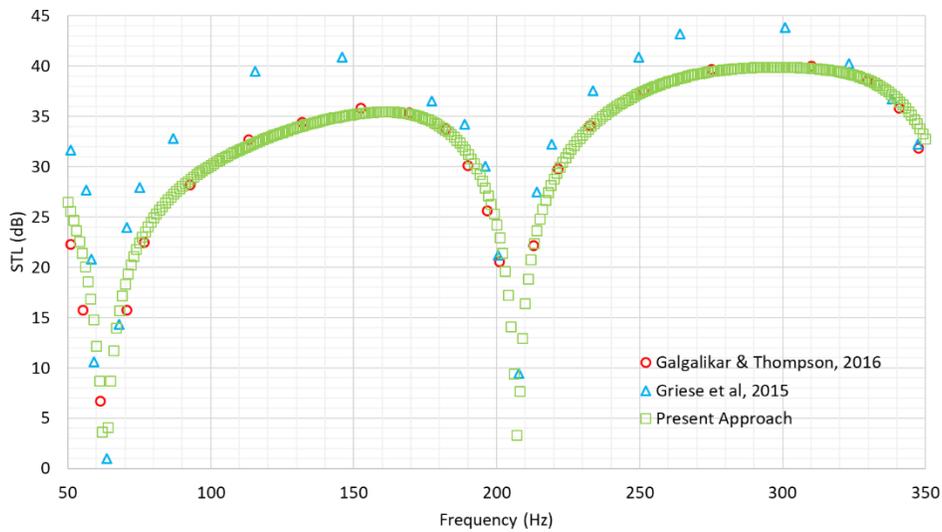

**Figure 6:** STL comparative study

Figure 7 shows the sound transmission loss versus frequency for four sandwich panels with different configurations of tubular cell cores. The geometrical characteristics of the cores utilized in these panels can be found in Table 1 along with the AUC parameter. Additionally, Table 2 provides information on the first six lowest natural frequencies of these sandwich panels. By analyzing the curves of various core configurations, we can assess the effectiveness of each design in reducing sound transmission. For instance, when considering frequencies between $150Hz$ and $250Hz$, sandwich panel sample 2 demonstrates a higher STL in comparison to the other sandwich panels shown in Figure 7. Although it may be straightforward to visually compare a small number of sandwich panel samples, as the number of sandwich panels increases, this task becomes more difficult.

**Table 1:** Geometrical parameters of the TCC and the $AUC$

| Sample | $M$ | $N$ | $t_c$ (mm) | $R_i$ (mm) | $L/H$ | $AUC$ ($dB.kHz$) |
|---|---|---|---|---|---|---|
| 1 | 7 | 50 | 1.36 | 8.65, 8.11, 9.14, 9.70, 8.04, 9.13, 8.25 | 11.6 | 12.269 |
| 2 | 9 | 48 | 1.20 | 6.29, 9.52, 10.27, 9.85, 10.01, 9.22, 7.05, 8.75, 10.00 | 8.4 | 12.299 |
| 3 | 5 | 38 | 0.86 | 12.31, 11.57, 12.98, 10.12, 9.10 | 13.6 | 10.944 |
| 4 | 6 | 58 | 1.58 | 5.28, 7.96, 6.92, 5.69, 7.15, 7.15 | 19.1 | 12.361 |

**Table 2:** The six lowest natural frequency ($Hz$) of the TCC sandwich panels

| Sample | 1st | 2nd | 3rd | 4th | 5th | 6th |
|---|---|---|---|---|---|---|
| 1 | 113.3 | 228.8 | 355.4 | 483.5 | 615.3 | 750.3 |
| 2 | 98.5 | 198.7 | 305.7 | 414.8 | 526.8 | 641.4 |
| 3 | 56.1 | 114.1 | 176.1 | 242.2 | 313.1 | 388.9 |
| 4 | 139.3 | 294.6 | 473.7 | 658.3 | 845.9 | 1034.8 |

Furthermore, as the frequency range increases, it becomes even more challenging to distinguish which sandwich panel has better acoustic performance. To simplify this process, the STL-Frequency plots can be quantified using the Area Under the Curve (AUC), as explained in Section 4. Utilizing the AUC enables an objective comparison of the performance of different sandwich panel samples. It allows us to identify the panel that exhibits the best overall acoustic performance within the desired frequency range.

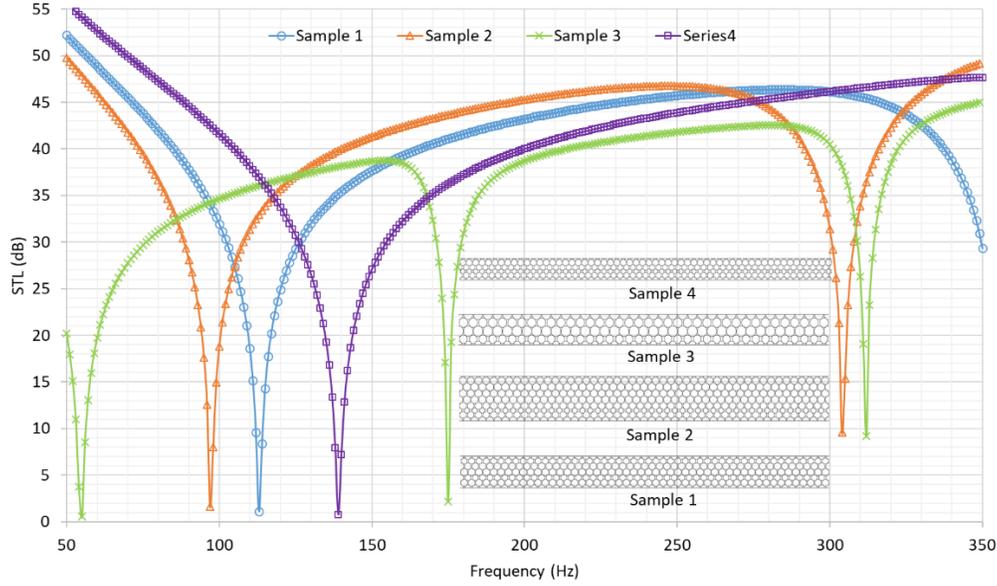

**Figure 7:** STL versus the frequency for TCC sandwich panels with different core configuration

In this research a total of 650 samples were generated, and among these samples, 500 were allocated for training the machine learning platform, while the remaining samples were reserved for testing the trained model. Table 3 presents the statistical measurements of the discrepancies between the $AUC$ predictions obtained using the GBDT-ML algorithm and those obtained based on the FEM analysis. These measurements include the mean, median, maximum, minimum, and standard deviation of the discrepancies, providing valuable insights into the accuracy and precision of the GBDT-ML algorithm's predictions. The results obtained from the test data indicate that the errors between the GBRT-ML algorithm and the finite element model fall within the range of -3.09% to 2.7% in 95% of the cases. This information offers an indication of the level of agreement between the two models, with the majority of the errors being relatively small. Furthermore, the high R-squared value of 0.96 suggests a strong relationship between the GBDT-ML algorithm and the FE model, indicating that the algorithm is well-suited for the data. In addition of testing the test data the GBDT-ML is used on the training data, which shows that the errors between the GBDT-ML algorithm and the finite element model are within a range of -0.29% and 0.26% in 95% of the cases. The low range of errors observed between the GBDT-ML algorithm and the finite element model in both the test and training data, as shown in Table 3, indicates that the proposed model has low variance and low bias. This suggests that the model has good generalization performance and is not much prone to over-fitting or underfitting the data.

**Table 3:** GBDT error statistical measures (%) in predicting the $AUC$

| Samples | Mean | Median | Max. | Min. | STD |
|---|---|---|---|---|---|
| **Test (150 Samples)** | 0.08 | 0.22 | 4.49 | -4.59 | 1.75 |
| **Train (500 Samples)** | 0.00 | 0.00 | 0.56 | -0.61 | 0.17 |

As a representative example, Table 4 demonstrates a side-by-side evaluation of AUC predictions acquired from the GBDT-ML algorithm and the FE model, focusing on a subset of 30 cases. The outcomes indicate a general alignment between the AUC values predicted by the GBDT-ML algorithm and those predicted by the FE model. By using this technique, engineers can quickly and accurately assess the acoustic performance of different sandwich panel designs and identify the most effective design for their specific application. The discrepancy, in Table 4, between the results obtained using the GBRT-ML algorithm and the FE model is calculated by:

$$Discrepancy = 100 \times \left(\frac{AUC_{\text{GBDT-ML}} - AUC_{FEM}}{AUC_{FEM}}\right) \qquad (22)$$

This research employs the NSGA-II algorithm to conduct a multi-objective optimization for the TCC sandwich panel. The objective is to maximize the AUC while simultaneously minimizing the panel's height and mass. To establish a reference design case, the design variables outlined in Section 4 are subject to an approximate averaging process. As a result, the reference design features $M = 6$, $N = 55$, $t_c = 1.0 \ mm$, and with the assumption of constant radius of $R = 7.5 \ mm$. This cell topology exhibits a resemblance, up to a certain extent, to the commonly used configurations found in existing literature [60], [61].

**Table 4:** Comparing AUC between GBDT and FEM Models

| M | N | $t_c$ (mm) | $R_i$ (mm) | $AUC_{FEM}$ (dB.kHz) | $AUC_{GBDT-ML}$ (dB.kHz) | Discrepancy (%) |
|---|---|---|---|---|---|---|
| 5 | 36 | 2.0 | 9.9, 13.0, 13.2, 13.5, 9.7 | 12.485 | 12.473 | -0.10 |
| 9 | 90 | 0.7 | 4.0, 5.0, 5.0, 3.3, 4.7, 3.5, 4.7, 3.7, 4.6 | 10.789 | 11.066 | 2.57 |
| 4 | 27 | 1.8 | 17.6, 11.5, 12.9, 16.9 | 11.402 | 11.390 | -0.11 |
| 8 | 78 | 0.7 | 6.0, 5.1, 5.1, 5.4, 6.0, 4.9, 4.8, 5.5 | 10.941 | 10.969 | 0.26 |
| 10 | 94 | 0.5 | 4.6, 4.4, 4.7, 4.4, 5.0, 4.8, 3.6, 4.6, 4.1, 5.2 | 11.171 | 11.113 | -0.52 |
| 6 | 43 | 2.0 | 11.2, 10.8, 8.0, 11.6, 8.5 | 13.008 | 13.009 | 0.01 |
| 7 | 49 | 1.3 | 10.1, 6.6, 8.1, 7.1, 10.0, 9.2, 8.6 | 11.843 | 12.033 | 1.60 |
| 2 | 12 | 2.9 | 34.8, 30.2 | 10.521 | 10.702 | 1.72 |
| 5 | 29 | 1.6 | 10.5, 11.1, 14.2, 10.4, 16.4 | 11.038 | 11.225 | 1.69 |
| 3 | 18 | 6.2 | 21.5, 25.0, 26.5 | 13.774 | 13.519 | -1.85 |
| 6 | 60 | 0.9 | 5.6, 5.8, 5.3, 8.1, 6.6, 6.2 | 10.635 | 10.771 | 1.28 |
| 10 | 72 | 1.2 | 6.3, 5.1, 6.9, 5.5, 6.9, 5.0, 5.0, 6.6, 6.5, 4.9 | 13.058 | 12.749 | -2.37 |
| 7 | 61 | 1.2 | 7.6, 7.7, 5.1, 6.6, 6.0, 7.6, 7.6 | 12.070 | 12.044 | -0.22 |
| 3 | 30 | 1.3 | 15.4, 10.1, 16.0 | 9.873 | 9.865 | -0.08 |
| 9 | 82 | 0.8 | 5.0, 5.1, 5.6, 4.7, 3.9, 4.8, 5.0, 4.6, 5.2 | 11.555 | 11.541 | -0.12 |
| 3 | 17 | 5.8 | 19.8, 28.8, 27.1 | 13.570 | 13.566 | -0.03 |
| 7 | 53 | 1.1 | 9.0, 5.7, 8.4, 9.0, 6.7, 8.1, 8.4 | 11.298 | 11.307 | 0.08 |
| 8 | 80 | 1.1 | 5.4, 5.6, 4.4, 6.1, 5.4, 6.2, 6.2, 5.5 | 12.488 | 12.481 | -0.06 |
| 5 | 46 | 2.6 | 10.3, 9.7, 9.1, 9.2, 8.1 | 13.204 | 13.004 | -1.51 |
| 10 | 98 | 0.4 | 5.1, 3.6, 3.7, 3.2, 4.7, 4.2, 3.7, 4.8, 3.5, 3.1 | 10.626 | 10.630 | 0.04 |
| 4 | 20 | 5.7 | 23.4, 19.2, 20.5, 21.1 | 14.393 | 14.242 | -1.05 |
| 7 | 70 | 0.3 | 5.2, 4.3, 4.4, 5.8, 6.0, 6.3, 6.3 | 10.552 | 10.544 | -0.08 |
| 9 | 75 | 0.8 | 5.6, 4.7, 5.8, 5.8, 4.5, 4.2, 6.2, 5.3, 6.4 | 11.283 | 11.283 | 0.00 |

Figure 8 presents the Pareto front, that illustrates the best possible trade-off designs between the optimized AUC, panel height, and panel mass. The Pareto front showcases the different possible design solutions that achieve varying levels of performance in terms of AUC, height, and mass. By analyzing Figure 8, one can gain insights into the trade-offs involved when attempting to improve the STL for desired range of frequency while simultaneously minimizing the height and mass of the panel. It allows for a visual understanding of the objective space and the potential compromises that need to be considered when optimizing the TCC sandwich panel.

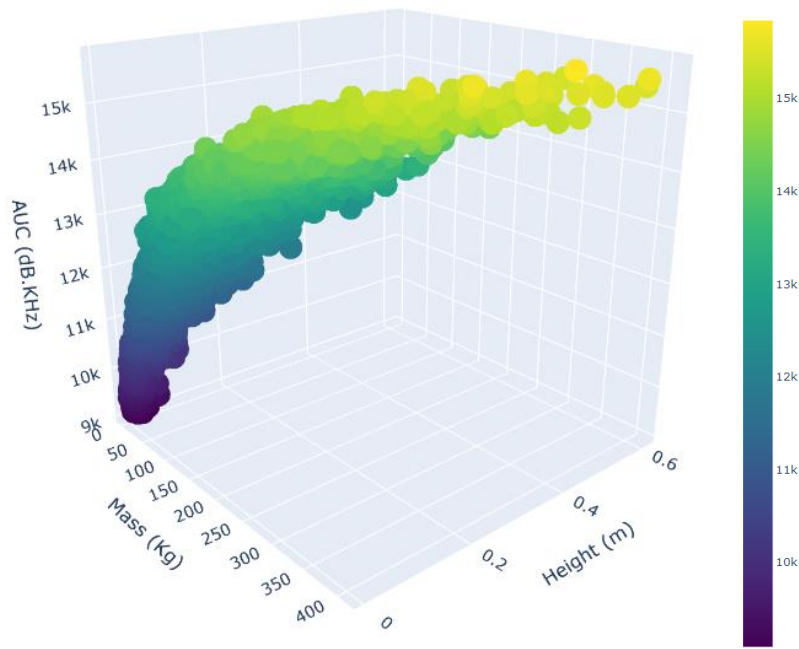

**Figure 8:** Pareto front (the best trade-off designs between the AUC, panel height and panel mass)

To minimize the impact of the inherent randomness in the NSGA-II algorithm, it is customary to execute the algorithm multiple times, 300 times in this case, and analyze the consolidated outcomes. This practice aims to identify solutions that are both robust and reliable. By running the algorithm repeatedly, we can obtain a more comprehensive understanding of the solution space and assess the consistency of the results. By filtering the non-dominated designs, we can pinpoint those that are the best trade-offs between the AUC, panel height, and panel mass. This approach enhances the confidence in the identified solutions and facilitates the selection of the most appropriate design for the TCC sandwich panel optimization. Table 5 presents three different optimized TCC sandwich panel samples (labeled as Samples A, B and C). Each sample has a unique combination of parameter values for $AUC$, mass, height, $M$, $N$, $t_c$, and $R_i$. These solutions represent potential design options that have been identified through the optimization process. The selection was made based on trade-offs between the optimized AUC, panel mass, and panel height. These criteria are important factors for the design of TCC sandwich panels. The samples in Table 5 demonstrate variations in the parameter values, which shows that the NSGA-II algorithm explored a range of design possibilities and identified diverse solutions with different trade-offs among the considered parameters.

**Table 5:** Comparing optimal trade-off configuration with reference design (RD) for the TCC sandwich panel

| Sample | $AUC$ $(dB.kHz)$ | Mass $(Kg)$ | Height $(cm)$ | $M$ | $N$ | $t_c$ $(mm)$ | $R_i$ $(mm)$ |
|---|---|---|---|---|---|---|---|
| A | 13.763 | 72.6 | 10.5 | 4 | 16 | 5.5 | 25.1, 22.4, 21.8, 27.3 |
| B | 14.480 | 119.0 | 18.34 | 5 | 14 | 5.9 | 28.4, 26.6, 32.2, 31.4, 23.9 |
| C | 11.330 | 13.4 | 4.6 | 3 | 28 | 2.3 | 10.9, 17.1, 13.3 |
| RD | 11.070 | 28.0 | 6.0 | 6 | 55 | 1.0 | 7.5, 7.5, 7.5, 7.5, 7.5, 7.5 |

Figure 9 serves as a visual representation of the STL-frequency curve, where the optimized TCC sandwich panels exhibit a notable increase in sound transmission loss compared to the reference panel. This indicates that the optimized panels are more effective at reducing the transmission of sound waves, resulting in better acoustic insulation. In addition, Table 5 is provided to give a comprehensive overview of the geometrical parameters and AUC values for the panels. These parameters include dimensions, thickness, and other key characteristics that contribute to the overall performance of the panels. The AUC values serve as a quantitative measure to assess the effectiveness of the panels in attenuating sound, with higher values indicating better acoustic performance. The comparison between sample C and reference design (RD) panel demonstrates the advantages of the optimized TCC sandwich panel. Despite having similar $AUC$ values, sample C stands out due to its significantly reduced mass (52% reduction) and height (23% reduction). This indicates that the optimization process successfully achieves comparable acoustic performance while offering the added benefits of lightweight and slender design, which can be advantageous in various applications such as aerospace or automotive industries.

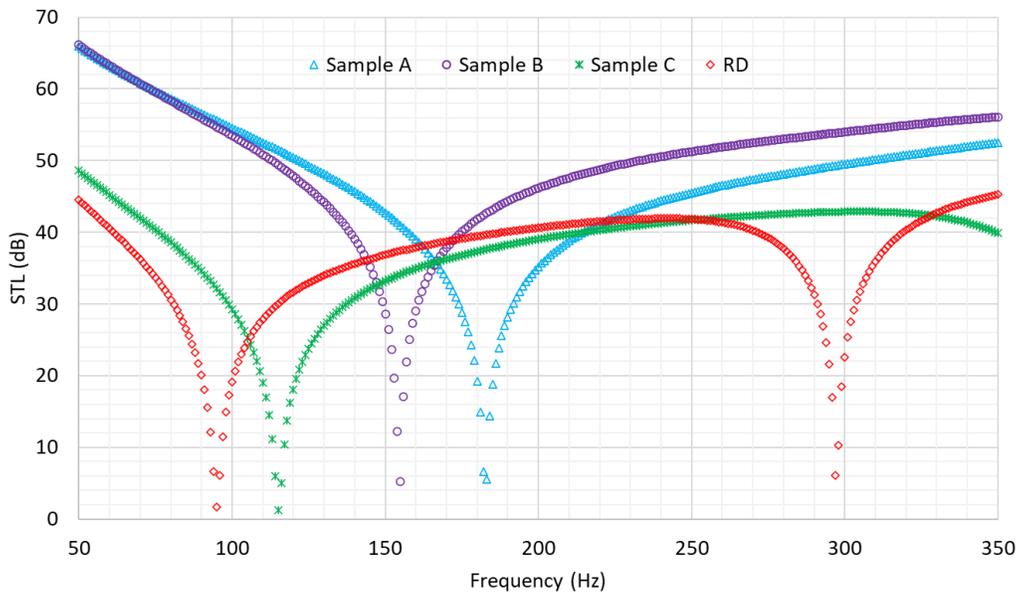

**Figure 9:** Contrasting the STL with the frequency curves of the optimized TCC sandwich panels (A, B, and C) in comparison to the reference design panel (RD).

Figure 10 presents a schematic view of the mode shapes and natural frequencies of the TCC sandwich panels mentioned in Table 5, allowing for a deeper understanding of their vibrational behavior. The mode shapes represent the distinctive patterns of panel vibration, while the natural frequencies indicate the frequencies at which resonance occurs. Samples A and B have only one natural frequency in the desired frequency range (50-350 Hz), while sample C and RD have more, which contributes to the fact that AUC of samples A and B is higher relative to others.

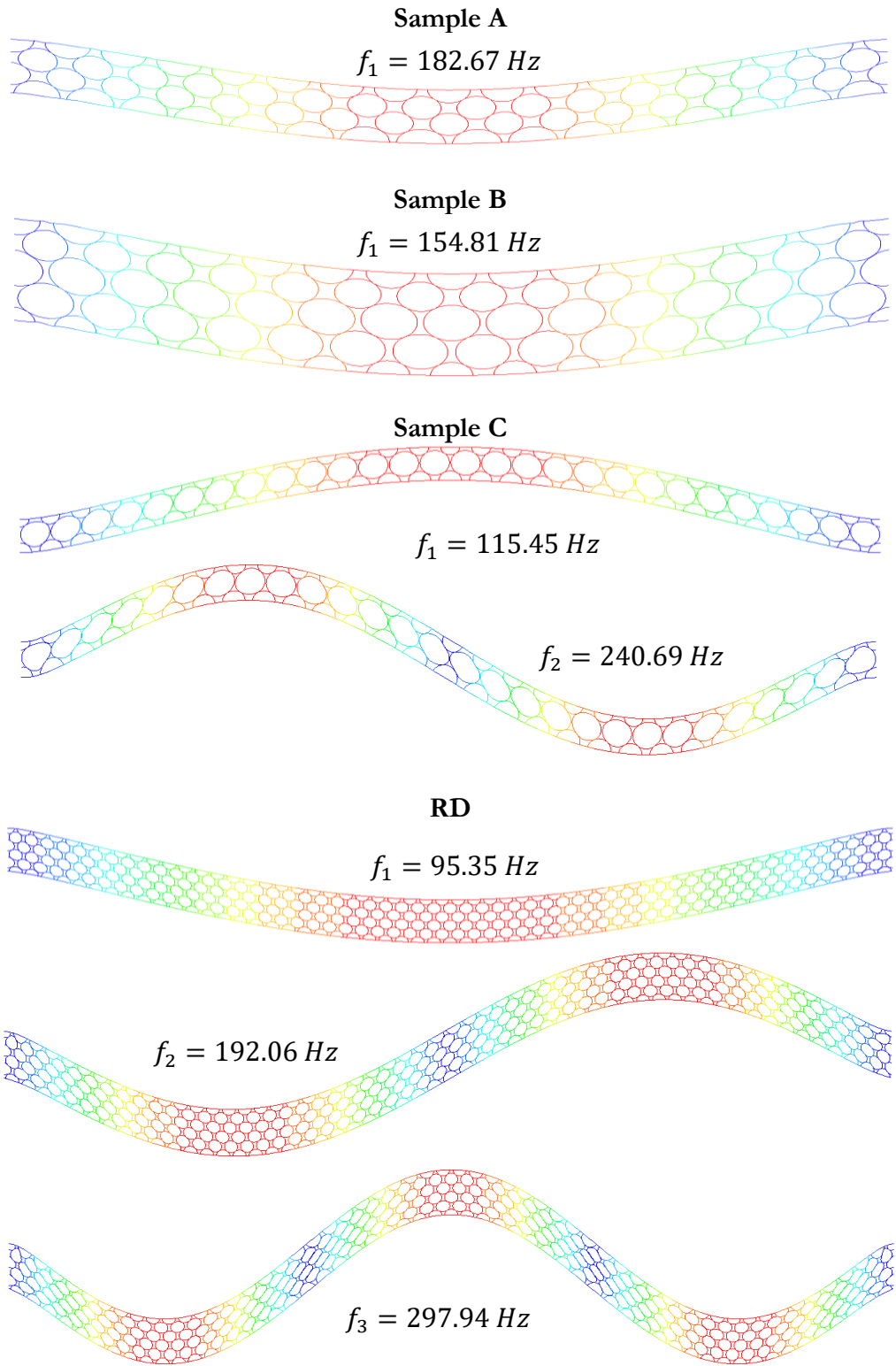

**Figure 10:** Schematic view of mode shape along with natural frequencies within the desired range of frequency ($50\ Hz$ to $350\ Hz$)

In the end, it could be said that the optimization of TCC sandwich panels results in superior acoustic performance compared to the reference panel. The combination of Figure 9 and Figure 10, along with the AUC analysis presented in Table 5, offers comprehensive insights into the improved sound transmission loss characteristics and geometrical parameters of the optimized panels. These findings provide valuable information for the design and development of acoustic insulation solutions in various industries.

## 7. Concluding Remarks

This research presents a theoretical approach to evaluate the sound transmission loss (STL) of sandwich panels with tubular cell core (TCC) and in-plane orientation. The Non-dominated Sorting Genetic Algorithm II (NSGA-II) is employed to optimize the core's geometric parameters, aiming to maximize the STL while minimizing the total mass and height of the panel. A machine learning technique based on Gradient Boosting Decision Trees (GBDT) is used to predict the STL of the panel, reducing computational costs. The accuracy and reliability of the proposed approach are validated through various analyses. Mesh convergence studies confirm the appropriate mesh configuration for accurate STL response calculations. Comparison with existing studies on honeycomb sandwich panels demonstrates favorable agreement, indicating the accuracy of the finite element solution procedure. The effectiveness of different tubular cell core configurations in reducing sound transmission is evaluated by analyzing the STL-frequency curves. The Area Under the Curve (AUC) parameter is introduced as a metric to objectively compare the acoustic performance of different panel designs within the desired frequency range. A machine learning model based on GBDT is trained and tested, showing a close alignment between the predicted AUC values and the finite element model results. The low range of errors indicates good generalization performance of the model. Optimization using the NSGA-II algorithm identifies several optimized TCC sandwich panel samples with improved acoustic performance. The Pareto front illustrates the trade-offs between

AUC, panel height, and mass, aiding in design decision-making. Multiple runs of the algorithm enhance robustness and reliability of the identified solutions. The optimized panels exhibit increased sound transmission loss compared to the reference panel, as depicted in the STL-frequency curves. The geometrical parameters and AUC values of the optimized panels are presented, demonstrating significant reductions in mass, more than 50%, and thickness, more than 20%, while maintaining comparable $AUC$ value. The mode shapes and natural frequencies analysis provides insights into the vibrational behavior of the TCC sandwich panels. The reduction of the number of natural frequencies within the desired frequency range and general increase of STL levels indicates effective sound attenuation, resulting in higher $AUC$ values for selected optimized panel samples (samples A and B). In conclusion, the optimization of TCC sandwich panels yields improved acoustic performance compared to the reference panel. The comprehensive analysis of STL, geometrical parameters, and vibrational behavior offers valuable information for the design and development of acoustic insulation solutions in various industries.

## 8. Acknowledgments

The authors acknowledge the financial support by Natural Sciences and Engineering Research Council of Canada (NSERC) [grant number RGPIN-2017-04509 and IRCPJ 550069-19].